\documentclass[twocolumn,twoside,slac_two]{revtex4}
\usepackage{graphicx}
\usepackage{fancyhdr}
\usepackage{natbib}
\begin{document}

\title{Uncovering low-level \emph{Fermi}/GBM emission using orbital background subtraction }

%

\author{G. Fitzpatrick}
\affiliation{School of Physics, University College Dublin, Belfield, Dublin 4 }

\author{V. Connaughton}
\affiliation{University of Alabama in Huntsville }

\author{S. McBreen, D. Tierney}
\affiliation{School of Physics, University College Dublin, Belfield, Dublin 4}

\author{On Behalf of the GBM Team}

\begin{abstract}
The secondary instrument onboard \emph{Fermi}, the Gamma-ray Burst Monitor (GBM) is an all sky monitor consisting of 14 scintillation detectors. When analysing transient events such as Gamma-Ray Bursts (GRBs) and Solar Flares (SFs) the background is usually modelled as a polynomial (order 0-4). However, for long events the background may vary more than can be accounted for with a simple polynomial. In these cases a more accurate knowledge of GBM's background rates is required. Here we present an alternative method of both determining the background and distinguishing low-level emission from the instrumental background. 
\end{abstract}

\maketitle

\thispagestyle{fancy}

\section{Motivation}

Launched into a low earth orbit in 2008, the \emph{Fermi} Gamma-Ray Space Telescope consists of two instruments, the Large Area Telescope (LAT)~\cite{atwood} which uses a pair-production system of detection and the Gamma-Ray Burst Monitor (GBM)~\cite{meegan} which consists of 14 scintillation detectors. Observing the entire unocculted sky, the GBM has an energy range of 8 keV-40 MeV which overlaps with the lower end of the LAT's range, 20 MeV-300 GeV.

\emph{Fermi} has an inclination of 26.5$^{\circ}$, an altitude of $\sim$565 km and a period of $\sim$ 96 minutes. The primary observation mode of is Sky Survey Mode, this optimises the sky coverage of the LAT whilst maintaining near uniform exposure. In this mode the satellite rocks about the zenith ($\pm$50$^{\circ}$ formerly $\pm$35$^{\circ}$) such that the entire sky is observed for $\sim$30 minutes every 2 orbits ($\sim$ 3 hours). In addition to this rocking the satellite pointing alternates between the northern and southern hemispheres each orbit. Due to the fact that \emph{Fermi}'s instruments are deactivated in the South Atlantic Anomaly (SAA), which is primarily in the southern hemisphere, there is an exposure differential of $\sim$15 \% between observations in the north and south hemispheres.

The GBM was optimised for the study of the prompt emission from GRBs, which is characterised by impulsive peaks with sharp rises, often highly structured, and easily distinguishable against instrumental backgrounds. The timescale on which this emission typically occurs is usually short enough that the background can be modelled as a polynomial of order 0-4. However, this method is not suited to resolving smoother long lived emission. 
\begin{figure*}[t]
\centering
\includegraphics[scale=0.32]{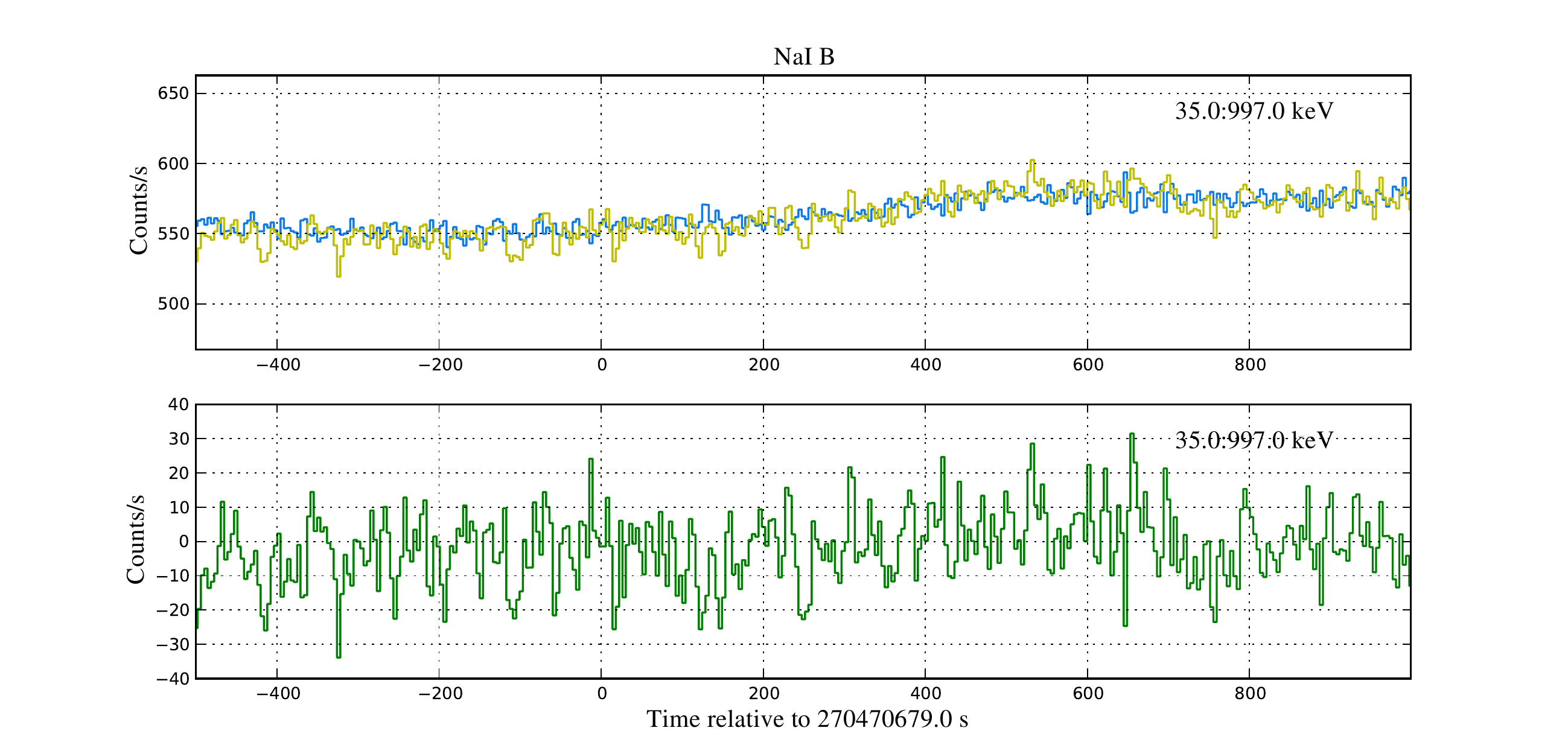}
\includegraphics[scale=0.32]{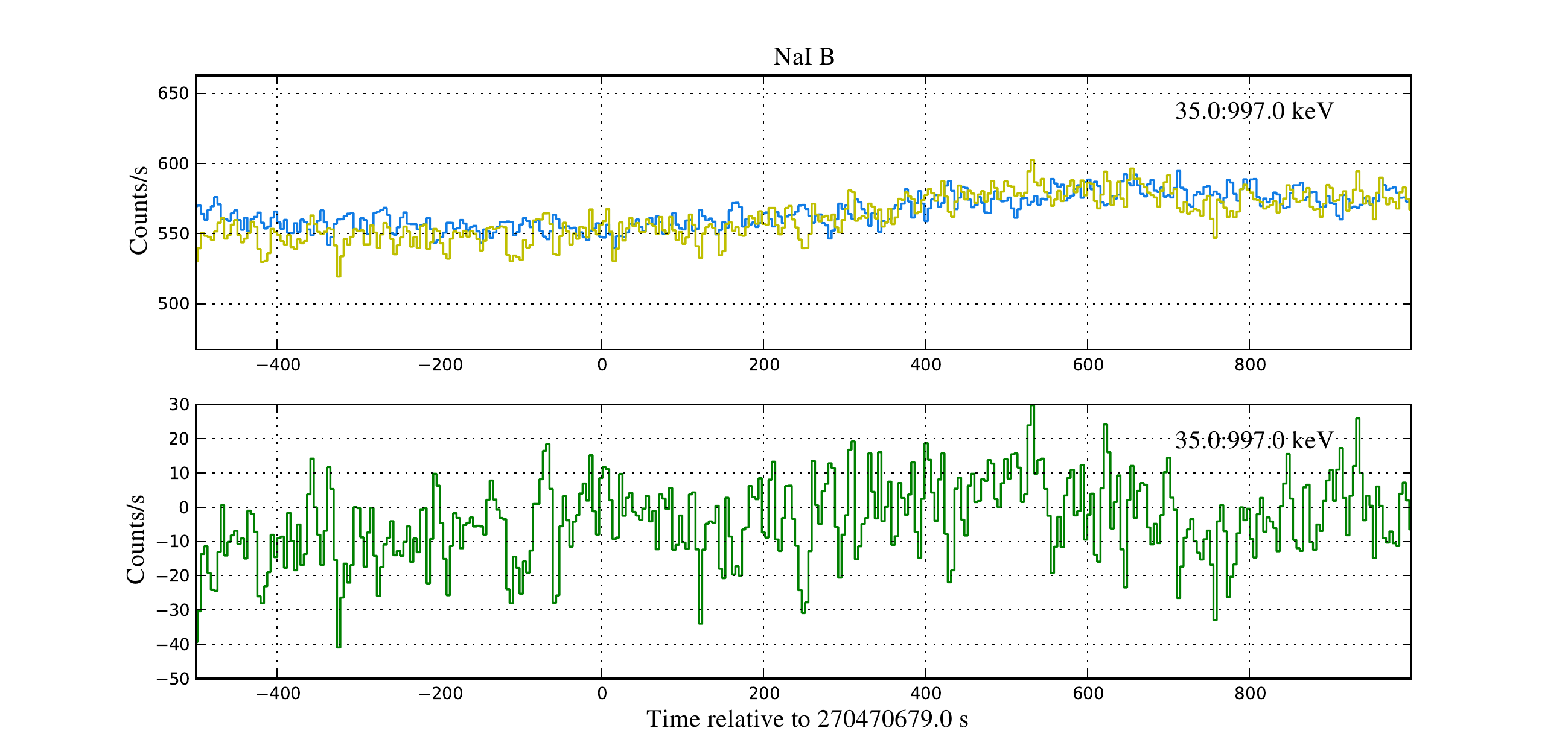}
\caption{Estimated background and source rates for detector NaI B ($\sim$35-1000 keV) determined using $\pm$14,16 (left panel) and $\pm$30 (right panel) orbits for region A. In each panel the upper plot is the source (yellow) and background (blue) rates and the lower is the residual rates. In both cases the estimated rates agree well with the source rates.} \label{lc}
\end{figure*}
The LAT has observed long lived emission on the order of ks from GRBs, which follows the prompt phase and usually decays as a power law (e.g. for GRB090926A following the prompt emission in the LAT ($\sim$25 s) the flux decayed as power low ($t^{-1.7}$)~\cite{ackerman11}).

To investigate whether such emission is detectable in GBM and also to confidently determine the background for solar flares, we have implemented a method for estimating the background which uses the rates from adjacent days, when the satellite is at the same geographical coordinates, to estimate the background at the time of interest. This project has been motivated by the work of Connaughton~\cite{val_tail}, who employed a similar technique with BATSE. 

\section{Method}
The orbit of \emph{Fermi} is such that it will be at approximately the same geographical coordinates every 15 orbits ($\sim$ 24 hrs). Therefore, it would seem that the background at time $T_0$ could be approximated by averaging the rates at times $T_0$ $\pm$15 orbits. However, this is not possible as the rocking angle of the spacecraft in sky survey mode is the same every 2 orbits and as a result detectors which are pointed towards the source at $T_0$ will not be pointed towards it at  $T_0$ $\pm$15 orbits. 

One solution to this is to use the rates from $T_0\pm$30 orbits. An alternative is to use the average of the rates from $T_0\pm$14 orbits and $T_0\pm$16 orbits to approximate the rates from $T_0$ $\pm$15 orbits.

An obvious limitation of this technique is that it cannot be employed to investigate GRBs for which the satellite accepted an Autonomous Repoint Request (ARR). When an ARR is trigged the telescope will slew so that the GBM calculated position is within the LAT FOV. A natural consequence of this is that the periodic pointing is interrupted for the duration of the ARR ($\sim$2 hours, formerly 5 hours).  This is unfortunate as it means that in general we cannot use our method to search for extended emission in GRBs with LAT detections where there is extended emission in the MeV-GeV range, as these will usually trigger an ARR. 

An additional issue is the passage of the satellite through the SAA which can cause elevated rates in the detectors due to activation, particularly in the BGO. The time spent in the SAA varies from orbit to orbit due to the precession of \emph{Fermi} ($\sim$52 days). This can lead to a systematic deviation between the estimated background rates and the source rates (see Section~\ref{saa_sec}).

\section{Blank Sky Tests} 
In order to test the validity of the method, a blank sky test was performed.  Between May 2009 and April 2011 four regions with triggerless periods of $\sim$4 days were selected and used as pseudo sources (see Table \ref{tab_details}). An additional criteria in the selection of the blank sky regions was that the regions offset $\pm$14,15,30 orbits did not contain and were not preceded by SAA passages.
\begin{table}[h]
\begin{center} 
\caption{Blank Sky test details: The four regions were selected with the criteria that they correspond to a period of $\sim$ 4 triggerless days and that the regions offset $\pm$14,15,30 orbits from the zero time did not contain SAA passages.}
\begin{tabular}{c c c c }
\hline
\textbf{Region} & \textbf{Date} & \textbf{Zero Time (MET)} &  \\
A & 09/07/28 & 270470679 & \\
B &10/08/23 &302922999 & \\
C &10/08/10  &324800173 & \\
D &11/04/18 &304200725 &\\
\hline
\end{tabular} 
\label{tab_details}
\end{center}
\end{table}

In order to determine which temporal selection provided the background most similar to the actual rate, the background was estimated from $T_0\pm$30 orbits and the average of $T_0\pm$14,16 orbits for all regions. The background was estimated for a duration of 3500 s using the continuous CSPEC data (128 energy channels and temporal resolution of 4.096 s). Sample lightcurves for detector NaI b for region A can be seen in Figure~\ref{lc}. 
\begin{figure*}[t]
\centering
\includegraphics[scale=0.3]{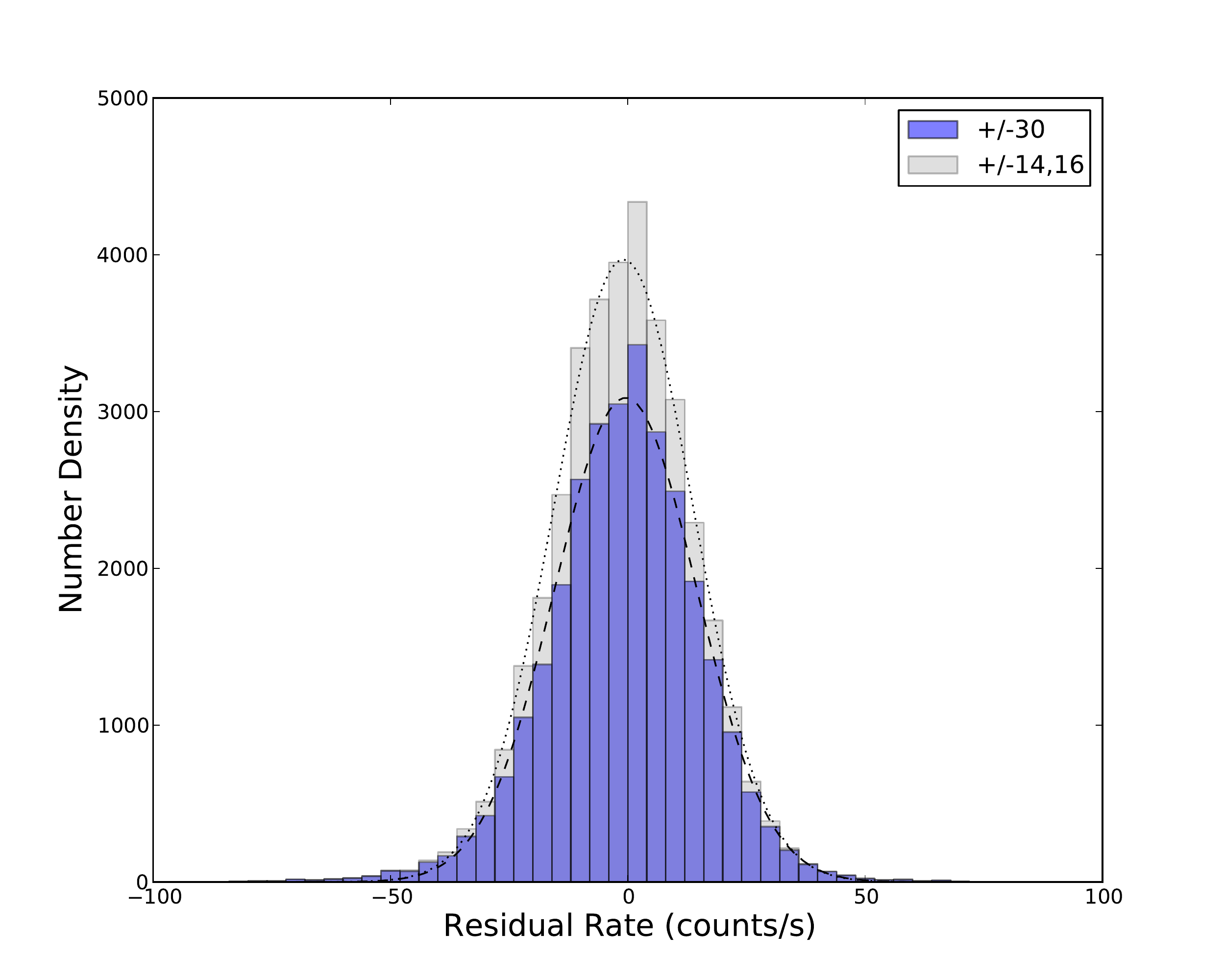}
\includegraphics[scale=0.3]{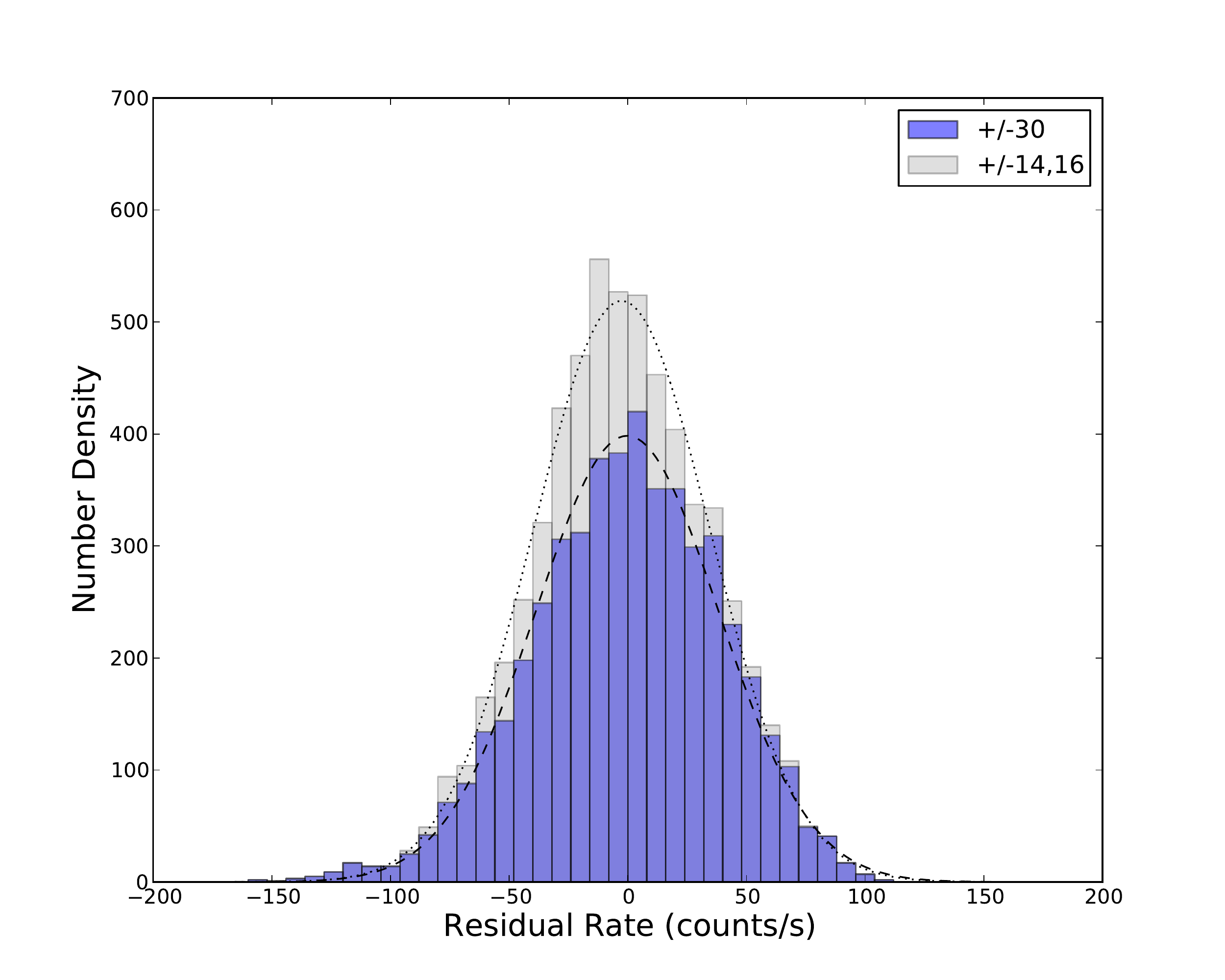}
\caption{Residual Rates for all NaI (left panel) and all BGO (right panel) detectors in all regions for $\pm$30 (blue) and $\pm$14,16 orbits (grey). } \label{hist}
\end{figure*}

\begin{figure*}[t]
\centering
\includegraphics[scale=0.32]{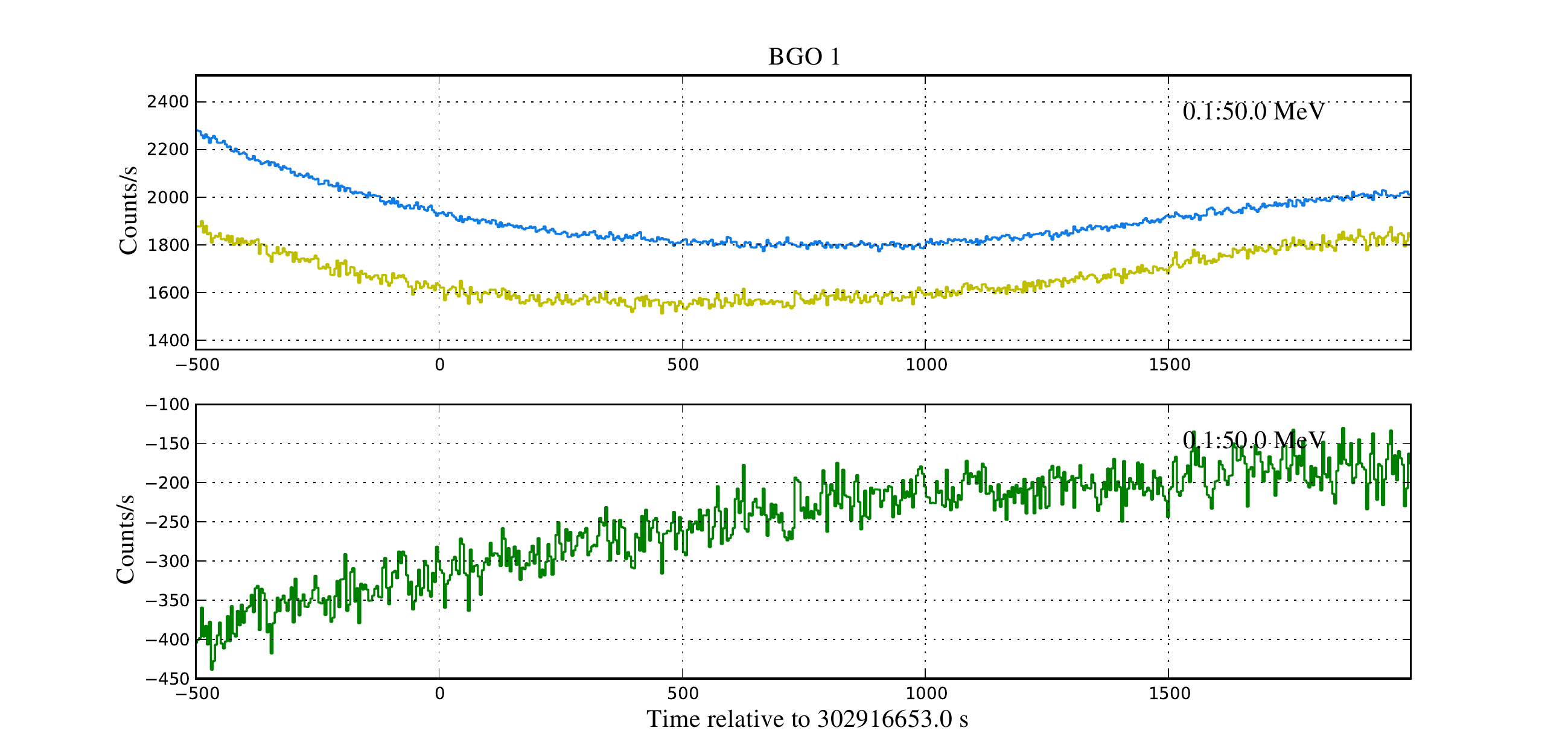}
\includegraphics[scale=0.32]{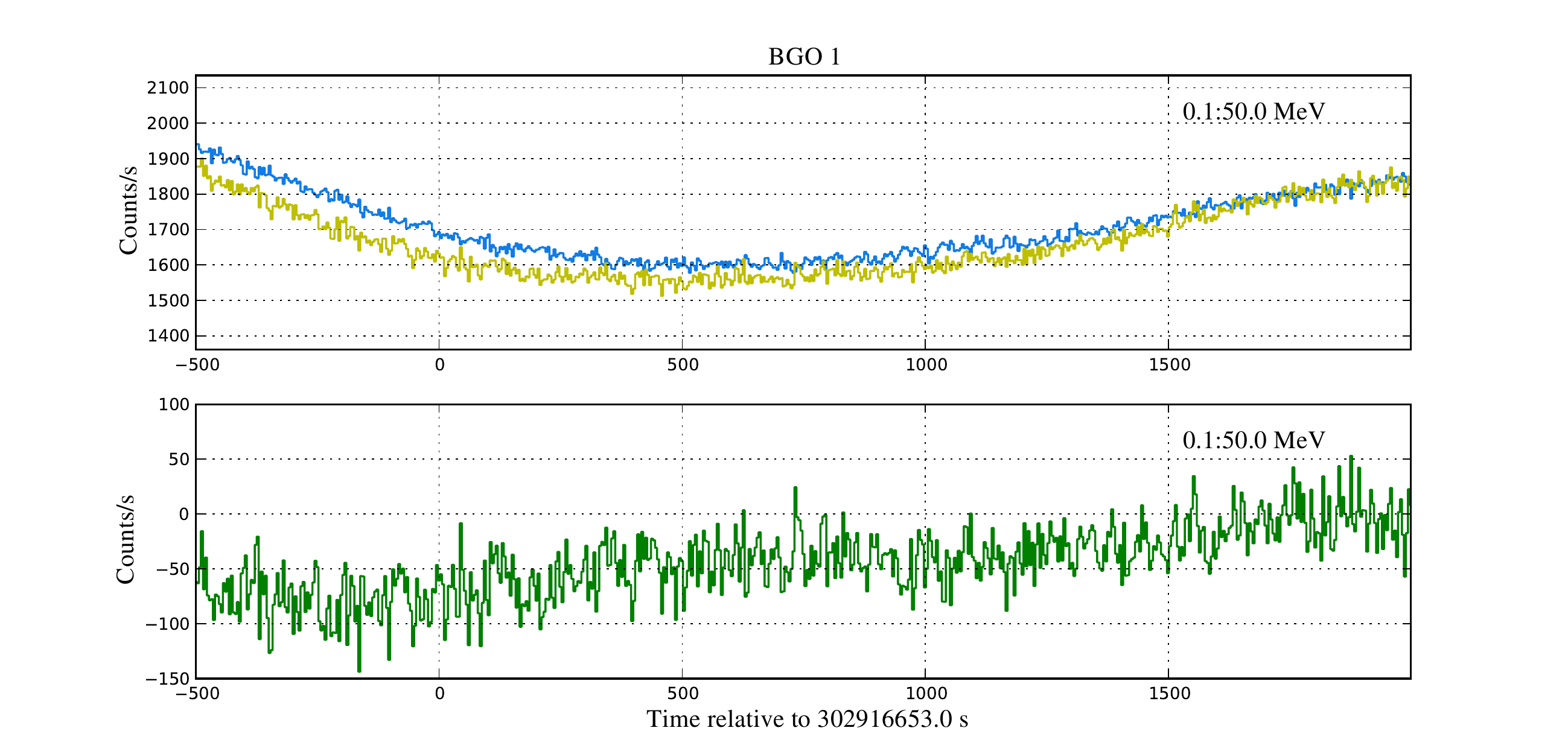}
\caption{Estimated background and source rates for detector B1 (0.1-50 MeV) determined using $\pm$14,16 (left panel) and $\pm$30 (right panel) orbits. In each panel the upper plot is the source (yellow) and background (blue) rates and the lower is the residual rates. The systematic offset between the source and background rates arises due to activation following SAA passage. This offset is much lessened for $\pm$30 orbits. } \label{activation_lc}
\end{figure*}

The systematic difference between the estimated rates and the actual rates was examined by histogramming the rate residuals from all the regions separately for the NaI and BGO. In order to limit the effect of interfering sources, the low energy channels of the NaI ($<$25 keV) were not included. Excluding also the overflow channel gives an energy range of $\sim$25-1000 keV. For the BGO only the first and overflow channels were discarded, corresponding to an effective energy range of $\sim$0.1-45 MeV. The results can be seen in Figure~\ref{hist}. For each offset ($\pm$30 and $\pm$14,16) the histogrammed residuals were fit with gaussians, the parameters of which can be seen in Table~\ref{fit_details}. The fit parameters are quite similar for $\pm$30 and $\pm$14,16 for both NaI and BGO which implies that both offsets are equally valid for estimating the background.

\begin{table}[t]
\begin{center} \scriptsize
\caption{Parameters from gaussian fit to histogrammed residual rates, all variables have units of counts/s. The fit parameters are quite similar for both $\pm$30 and $\pm$14,16 orbits. }
\begin{tabular}{c c c c c cc}
\hline 
& & $\pm$30 &  & &$\pm$14,16& \\
&\textbf{Amplitude} & \textbf{Mean} & \textbf{Sigma} &\textbf{Amplitude} & \textbf{Mean} & \textbf{Sigma} \\
NaI &3087 & -0.45 & 15 & 3970 & -0.9 & 14.6 \\
BGO &399 & -0.26& 38 & 518 & -2.5 & 37 \\
\hline
\end{tabular} 
\label{fit_details}
\end{center}
\end{table}

\subsection{Effect of SAA passage}\label{saa_sec}
Inspection of the residual lightcurves for a region preceded by an SAA passage showed that the background estimated from $\pm$30 orbits more closely matched the observed rates following an SAA passage, than that derived from $\pm$14,16. This is demonstrated in Figure~\ref{activation_lc}. This is to be expected as the time spent in the SAA for the $\pm$30 orbit offset  will be a closer match for the passage time for the source region. For regions where the effect of the SAA passage is negligible the rates from $\pm$30 and $\pm$14,16 orbits closely match the observed rates. This is beneficial as for these regions they can be used interchangeably, allowing the background to be estimated with $\pm$30 orbits if there is an interfering source in $\pm$14,16 orbits and vice versa.

\section{Conclusion}
Attempting to study long-lived or non-impulsive smooth emission in a background limited instrument in GBM is challenging. In order to do so a method using the rates from adjacent days to estimate the background has been developed. For the 4 triggerless periods studied this method generates a background which closely matches the source rates. Our study has shown that the rates from $\pm$30 and $\pm$14,16 orbits can be used interchangeably to estimate the background at the time of interest unless there is an SAA passage exit close to the time of interest, in which case the rates from $\pm$30 should be used. 

\subsection{Acknowledgments} \label{Ack}

\begin{acknowledgments}
This work has been suppourted by a Marie Curie European Reintegration Grant within the 7th Program under contract number PERG04-GA-2008-239176 and by the European Space Agency/Enterprise Ireland.
\end{acknowledgments}

\end{document}